\documentclass[%
 aip,
 jmp,%
 amsmath,amssymb,
 reprint,%
]{revtex4-2}

\usepackage{graphicx}
\usepackage{dcolumn}
\usepackage{bm}

\begin{document}

\preprint{AIP/123-QED}

\title[Confined colloidal droplets dry to form  circular mazes]{Confined colloidal droplets dry to form  circular mazes}

\author{Ilaria Beechey-Newman}
\affiliation{PoreLab, Department of Physics, Norwegian University of Science and Technology.}
\author{Natalya Kizilova}
\affiliation{ 
Institute of Aeronautics and Applied Mechanics, Warsaw University of Technology}
\affiliation{Department of Applied Mathematics, V.N. Karazin Kharkiv National University.}
\author{Andreas Andersen Henning}
\affiliation{PoreLab, Department of Physics, Norwegian University of Science and Technology.}
\author{Eirik Grude Flekkøy}
\affiliation{PoreLab, The Njord Centre, Department of Physics, University of Oslo}
\affiliation{PoreLab, Department of Chemistry, Norwegian University of Science and Technology.}
\author{Erika Eiser}
\affiliation{PoreLab, Department of Physics, Norwegian University of Science and Technology.}
\affiliation{Department of Physics, University of Cambridge.}

\date{28th February 2025}

\begin{abstract}
During drying, particle-laden sessile droplets will leave so-called coffee-stain rings behind. 
This phenomenon is well-known and well-understood (Deegan et al., Nature \textbf{389},827–829 (1997)).
Here we show that when particle-laden droplets confined in a slit are allowed to evaporate very slowly, they do not deposit coffee rings, but form a surprisingly intricate, circular maze-like pattern. 
We present experiments that illustrate this pattern formation and discuss the factors that determine when such patterns can form.
We are not aware of reports of natural examples of the formation of such beautiful patterns under confinement, although it seems likely that they exist.
\end{abstract}

                             
\maketitle

\section{Introduction}
Patterns exist everywhere in the natural world. 
Striking examples are Romanesco broccoli\,\cite{brocolli2021}, snail shells, or the arrangements of sunflower seeds\,\cite{ridley1982packing}. 
Similarly, animals exhibit a multitude of striking geometrical patterns, and the human brain has evolved to recognise such patterns~\cite{recongition1976}. 
Of course, we do not only enjoy observing intricate natural patterns, we also want to understand the physical mechanisms by which they form. 
An early example of the physics behind pattern formation is the work of Turing on the patterns generated by reaction-diffusion processes~\cite{turing1952,turing1990chemical}.

However, patterns need not be intricate to be interesting from a physics perspective. 
A case in point is the mundane ``coffee-stain'' effect that is observed when a sessile droplet containing a particulate dispersion ({\em e.g.}\ coffee) dries on a flat, partially wetting surface.
The formation of the resulting circular coffee stain was studied and explained in a number of papers by Deegan {\em et al.}~\cite{deegan1997capillary,Deegan2000Pattern}.
These papers argue that the accumulation of a ring of particles is primarily driven by the hydrodynamic flows caused by the high rate of evaporation at the rim of the sessile droplet.
For many practical applications, such as ink-jet printing~\cite{das2021chemically}, it is important to control the shape and uniformity of the solid deposits, which depend strongly on the size of the drops and their evaporation rate \cite{Kaplan2015}.
Depending on the circumstances,  drying sessile droplets may form single or multiple circular rings, uniform deposits, or other simple patterns.
Yet, they typically do not exhibit the intricate patterns caused by, say, diffusion-limited aggregation processes~\cite{thampi2023drying}.

\begin{figure}
    \centering
    \includegraphics[scale=0.45]{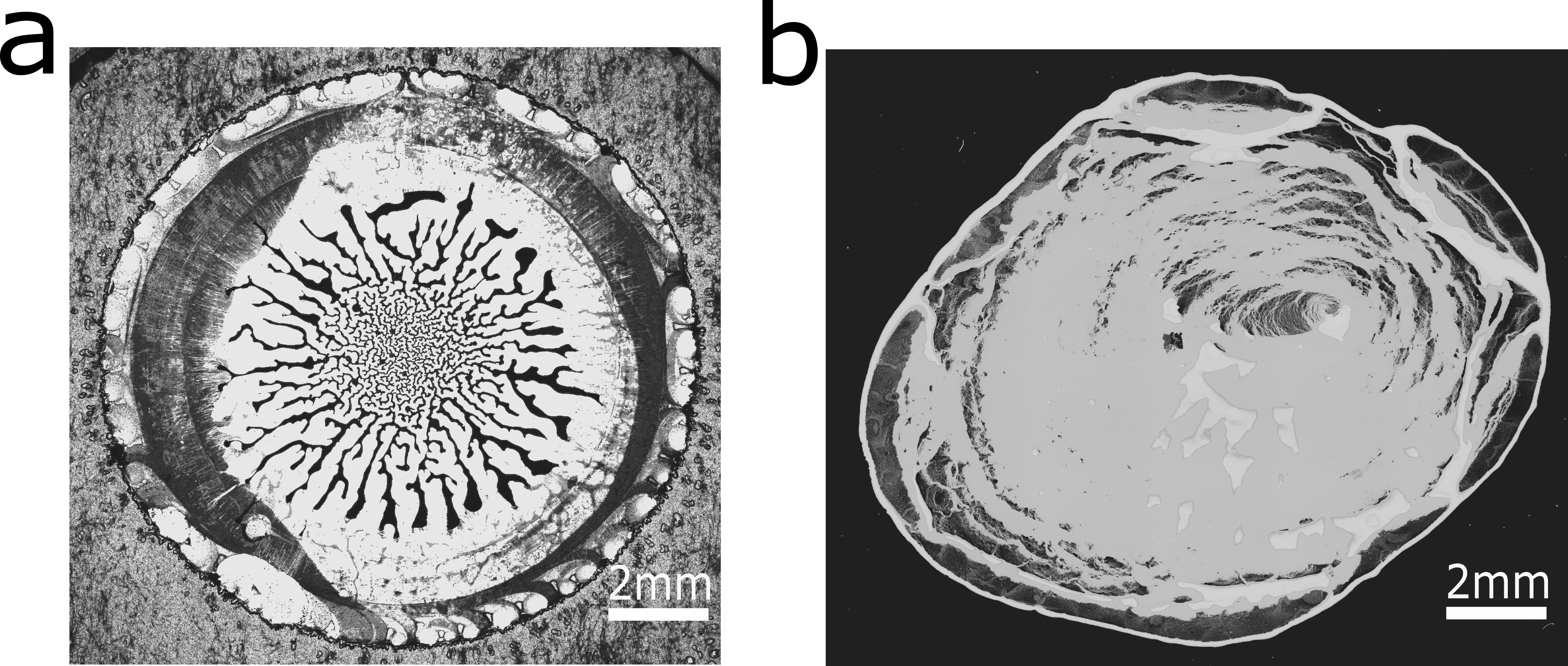}
    \caption{Drying pattern produced by a confined droplet of an aqueous suspension containing 1\,wt\% of charge-stabilized colloids with a diameter of 1.8\,$\mu$m, compared to an equivalent, unconfined droplet. The bright-field microscopy images show (a) the sediment deposited by a slowly dried droplet in cylindrical confinement,  and (b) the sediment deposited by a sessile droplet that has evaporated in dry air.}
    \label{fig:1}
\end{figure}

Here we show that vertical confinement of drying, particle-laden droplets can lead to very different (and aesthetically pleasing) patterns. 
In particular, we consider the practically important case of cylindrical, particle-laden droplets bridging the top and bottom of the confining slit, under conditions where the evaporation rate is throttled to be very low. 
Under these conditions, the shrinking droplet undergoes a sequence of instabilities resulting in the formation of a spherical ``maze'', with a pattern that gets finer towards the centre (Fig.\,\ref{fig:1}).
 
As we argue below, this pattern formation is possible under conditions where the rim of the droplet is not pinned to the surface. 
At some point, the accumulation of colloids advected to the side walls of the cylindrical droplet causes the effective surface tension to change sign, leading to a fingering instability.
The tips of these fingers are arrested, but the invaginations in between (the ``air fingers'') can grow, and split, leading to ever finer channels.

\section{Experiments}

\subsection{Drying process}

\begin{figure}
    \centering
    \includegraphics[scale=0.36]{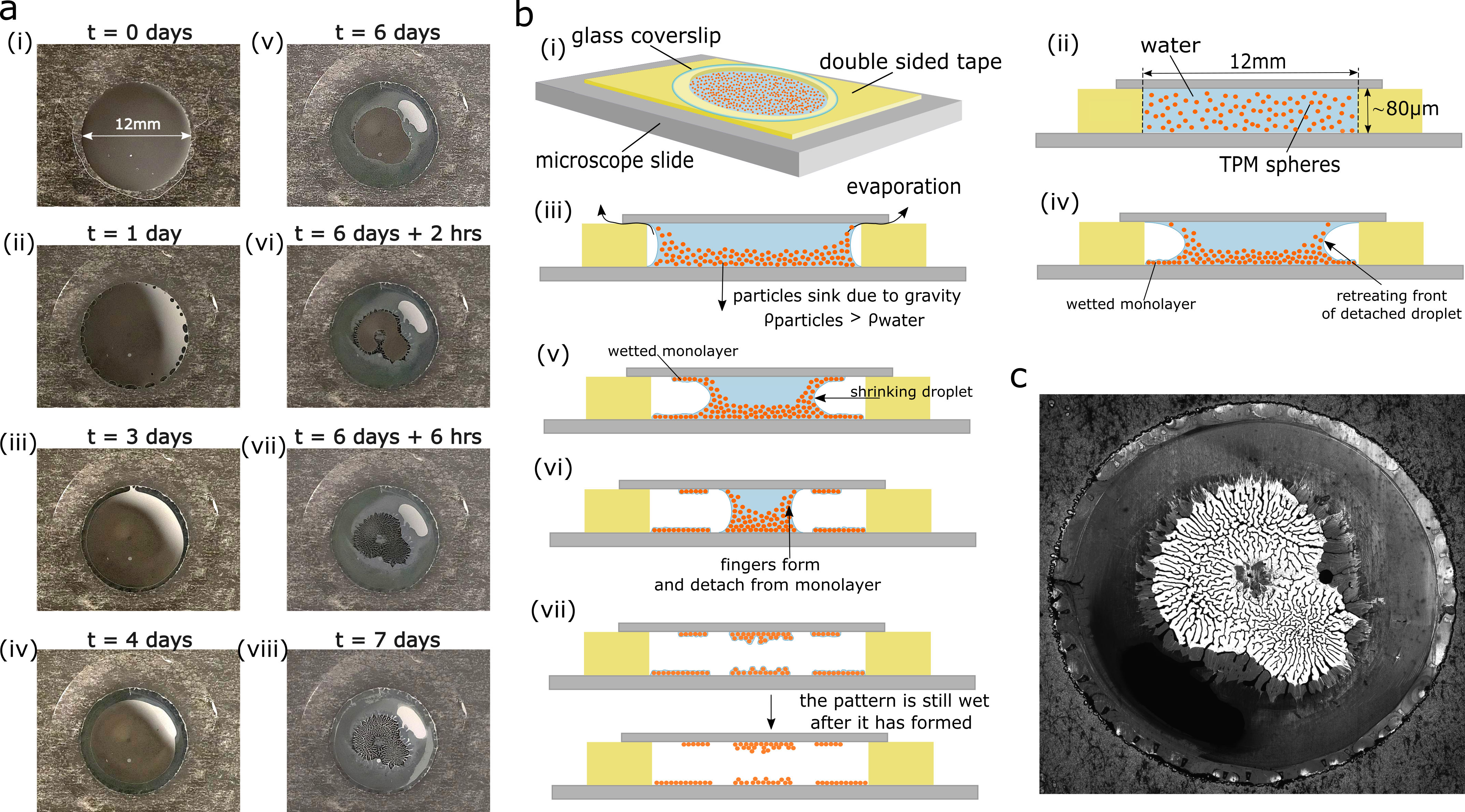}
    \caption{Different stages of a drying droplet confined and quasi-sealed between two glass surfaces. (a) Photographs of the sample from time-lapse movie V1, taken during 7 days of drying. (b) Schematic of (i) the sample geometry, (ii) a cross-section of the sample at $t = 0$, (iii) the bubble formation due to evaporation of water across the gap between the sticky tape and the cover slip, (iv) the droplet detachment from the edge and the beginning of a monolayer deposition on the bottom surface. (v) Subsequently, the sedimented colloids are pushed up the air-water interface, driving a monolayer deposition on the cover slip as well. (vi) The monolayer deposition stops abruptly, followed by finger formation, through which the fingers remain connected between the top and bottom surfaces. (vii) The final pattern of fingers is formed and the liquid bridge between the surfaces breaks: similar colloidal patterns remain on both surfaces but are still wet. (c) Microscopy image of the fully dried sample.}
    \label{fig:2}
\end{figure}

Fig.\,\ref{fig:2}a shows a sequence of photographs (taken from video V1) of an evaporating cylindrical droplet containing a dispersion of spherical TPM colloids (diameter 1.8\,$\mu$m -  see Methods section), bounded by glass surfaces on top and bottom, and confined in a quasi-sealed cylindrical cavity. Fig.\,\ref{fig:2}b shows a cross-sectional schematic drawing of the droplet evolution, and Fig.\,\ref{fig:2}c shows the pattern that remains once the droplet has dried. 

The cylindrical cavity containing the dispersion was created by punching a hole of 12\,mm in diameter in a piece of double-sided sticky tape that was pressed onto a clean microscope slide. 
After placing $\sim$30\,$\mu$L  of a 1\,wt\% colloidal suspension in the cylindrical cavity, a circular coverslip was carefully placed on top of it. 
This procedure resulted in a slight overfilling of the cylindrical cell, which had a height of $\approx 80\,\mu$m. 
The cover slip was not firmly pressed onto the sticky sealing tape, thus allowing for very slow evaporation of water vapor out of the cell and slow permeation of air into the cell (see Fig.\,S1 and video V2). 

Typical drying times were $8\pm 2$ days: if the evaporation had not been throttled, the same droplet would evaporate in a matter of minutes.
This point is important because it implies that the escape of water vapor through the cell walls is the rate-limiting step in the evaporation.
This observation implies that the water vapor pressure throughout the cylindrical cell is very nearly constant at a value that is just a few percent below the vapor pressure of water at the ambient temperature and that the evaporative mass flux of water is effectively constant, except at the very end of the drying process. 
The evaporation of the droplet was observed to proceed in three distinct stages. 
In the first stage (\textbf{stage 1}), starting at the moment of the sample preparation, we observed the slow formation and growth of air bubbles at the edge of the cell (Fig.\,\ref{fig:2}a(ii)). 
After about 4 days, these bubbles had grown enough to merge to form an annulus, enclosing a cylindrical droplet of the TPM suspension. 
The sketches in Fig.\,\ref{fig:2}b(ii-iv) present a cross-sectional view of this capillary bridge. 
Occasionally, we observed that the initially even distribution of colloids, seen as homogeneous whitish scattering light, photographed in reflection, becomes skewed (see Fig.\,\ref{fig:2}a, where more colloids can be seen in the top right of the sample). 
The formation of such off-center colloidal patches only happens if the sample is not held perfectly horizontal: after the TPM particles sediment, which typically happens within hours, the sedimented colloids can still diffuse, causing the sediment to accumulate preferentially on one side of the cell.  
This asymmetric accumulation of colloids in the droplet does not change the qualitative features of the subsequent evolution of the drying pattern.

Interestingly, microscopy experiments show that when the detached droplet shrinks (\textbf{stage 2}), a colloidal monolayer is deposited on the bottom surface, {\em and}, to a lesser extent, on the top surface.
We note that the barometric height of the TPM colloids, $h_B=\Delta mg/k_BT$, is only slightly larger than the colloidal diameter ($k_B$ is Boltzmann's constant, $T$ the absolute temperature, $ \Delta m$ the mass difference between the colloid and displaced water, and $g$ the gravitational acceleration).
The deposition of colloids on the top surface is therefore surprising, as the cell height is some 50 times larger than $h_B$.
The reason that the colloids can reach the top surface is that the evaporation-driven hydrodynamic flow to the droplet-air interface is strong enough (high enough P\'{e}clet number) to cause a flow-driven colloidal pressure near the interface that is large enough to cause the accumulation of the layer to expand in the vertical direction. 
In the sample shown in Fig.\,\ref{fig:2}a, the colloidal monolayer deposition on the bottom and top surfaces of the cell took place from day 4 to day 6.
Colloidal monolayer deposition by the retracting contact line of a colloid-laden droplet has been well studied \cite{denkov1993two,dimitrov1996continuous} and has been exploited to grow defect-free crystalline colloidal layers for photonic applications\,\cite{jiang1999single}.
There is a sample-to-sample variation in the density of the deposited colloidal layers, but they always appear after the droplet has detached from the walls of the cell.
Microscopy imaging of fluorescently-labeled TPM particles (see Fig.\,S2 and video V3) confirms the deposition of the TPM colloids on both the bottom and the top surfaces. 

The width of the accumulation region of TPM particles advected to the air-water interface is 20-30\,$\mu$m, which rules out that this layer is due to the ``Pickering'' mechanism where colloidal particles are embedded {\em in} the interface, as seen for particle-laden oil-in-water droplets\,\cite{sacanna2007thermodynamically,van2017preparation}.

\begin{figure}
    \centering
    \includegraphics[scale=0.325]{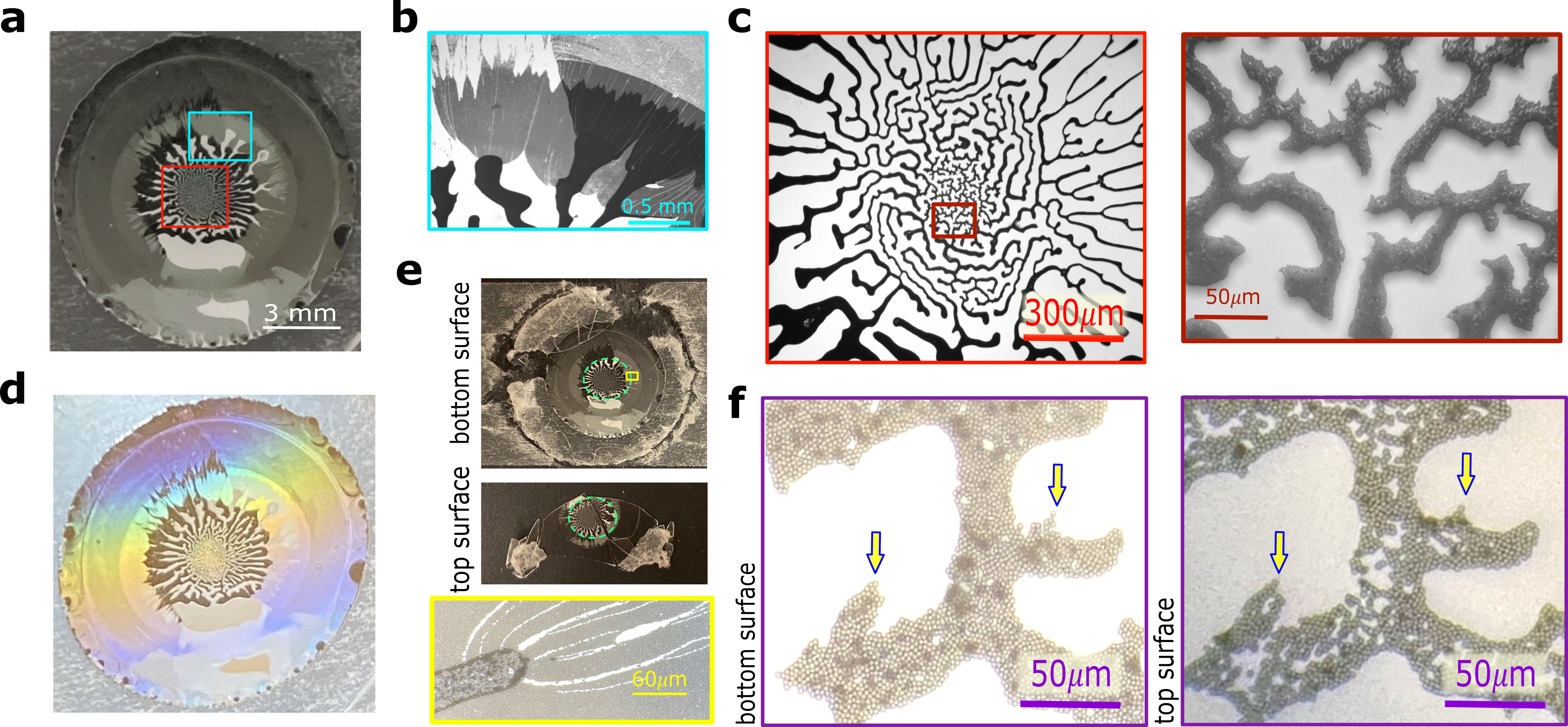}
    \caption{Photographs and bright-field microscopy images of a fully dried sample. (a) Photograph taken in reflection. Bright-field microscope images obtained of the (b) cyan and (c) red encased regions shown in (a), taken in transmission. (b) shows the monolayer-to-finger transition, while the images in (c) illustrate the abrupt change in length scales and orientations of the fingers, taken with increasing magnification. (d) Photograph of the sample taken in transmission, using the white light source of the microscope. (e) Photographs of the opened-up sample show similar dried patterns on the bottom and top slide (green-dashed circles), while the yellow rectangle is a zoom-in on one of the monolayer-to-finger transition regions on the bottom surface. (f) High magnification micrographs of the mirror-imaged colloidal arrangements in the center of the sample. The arrows show mirrored features.}
    \label{fig:3}
\end{figure}

\textbf{Stage 3} is marked by a dramatic change in the evolution of the drying droplet: the monolayer deposition stops abruptly, and the roughly cylindrical symmetry of the droplet is broken, as can be seen in Figs.\,\ref{fig:2}\,a(vi-viii).
These figures show the formation of fine, white ``fingers'' containing the colloids and black, colloid-free regions outside the shrinking droplet.
This finger formation is rapid in comparison with the first two stages and is completed after around 4-5 hours in this sample.  
However, the drying process is not yet finished at this point the , as the 'white' fingers containing the colloids are still wet, and colloidal motion can be observed inside them - even at the interface.
Moreover, the droplet still connects the top and bottom surfaces of the cell as depicted in Fig.\,\ref{fig:2}\,b(vi) and Fig.\,\ref{fig:4}\,d. 
Confocal microscopy images, shown in Fig.\,S2, demonstrate that the entire labyrinth structure and the deposited monolayers appear to be still wet after the fingers have formed.  
The water bridge between the top and bottom only ruptures after $t = 7$ days. 
After the liquid bridge between the top and bottom had ruptured, we could open up the cell and image the patterns left on the opposing surfaces (see Fig.\,\ref{fig:3}e).  
These images show that the pattern on the top surface and the mirror image of the bottom pattern are indeed very similar, as is to be expected if they resulted from the drying of the same droplet.

Several factors may distort the radial symmetry of the drying patterns.
For instance, trapped bubbles (as in the sample shown in Fig.\,\ref{fig:2}) may distort the pattern, and, as previously discussed, even a slight lack of horizontal alignment can distort the pattern.  
A gallery of drying patterns of analogous samples is presented in Fig.\,S4, demonstrating the reproducibility of this process.
Video V1 shows a time-lapse movie of the drying process depicted in Fig.\,\ref{fig:2}.

During stage \textbf{stage 3} there is a sudden change in the length scale and/or orientation of the fingers, as is demonstrated in Fig.\,\ref{fig:3}. 
In that figure,  we present another fully dried sample photographed both in reflection (Fig.\,\ref{fig:3}a) and transmission mode. 
The latter was taken using the white-light source of the microscope: the iridescent colors in Fig.\,\ref{fig:3}d stem from the colloidal monolayers that diffract the transmitted light. 
Fig.\,\ref{fig:3}b focuses on the transition from the monolayers to fingers, while Fig.\,\ref{fig:3}c focuses on the central finger-pattern and a further zoomed-in image of the finer structure at the center, where we can see individual colloids. 

The yellow-framed microscope image zooms in on a finger in contact with the monolayer on the bottom surface. 
The white streaks in the monolayer that seem to radiate from the fingers are cracks that appear in the final drying stage. 
These are associated with a volume reduction during drying\,\cite{li2005improved}.

\subsection{Interpretation}

As is clear from the above discussion, the pattern formation that we observe depends on the synergy of several factors:
the droplet must be confined, it must evaporate very slowly, the contact lines (top and bottom) should not be pinned (at least not initially) and the diffusion of the colloids should be fast enough to inhibit the formation of a ``thick'' coffee ring that would pin the radius of the droplet. 
These are just some of the factors that have to be ``just right'' (others include the nature of the solvent and the shape and concentration of the colloids).
Below, we provide a tentative explanation of how the different factors align to generate the intricate drying patterns that we observe, as presented in Fig.\,\ref{fig:3}.

\begin{figure}
    \centering
    \includegraphics[scale=0.6]{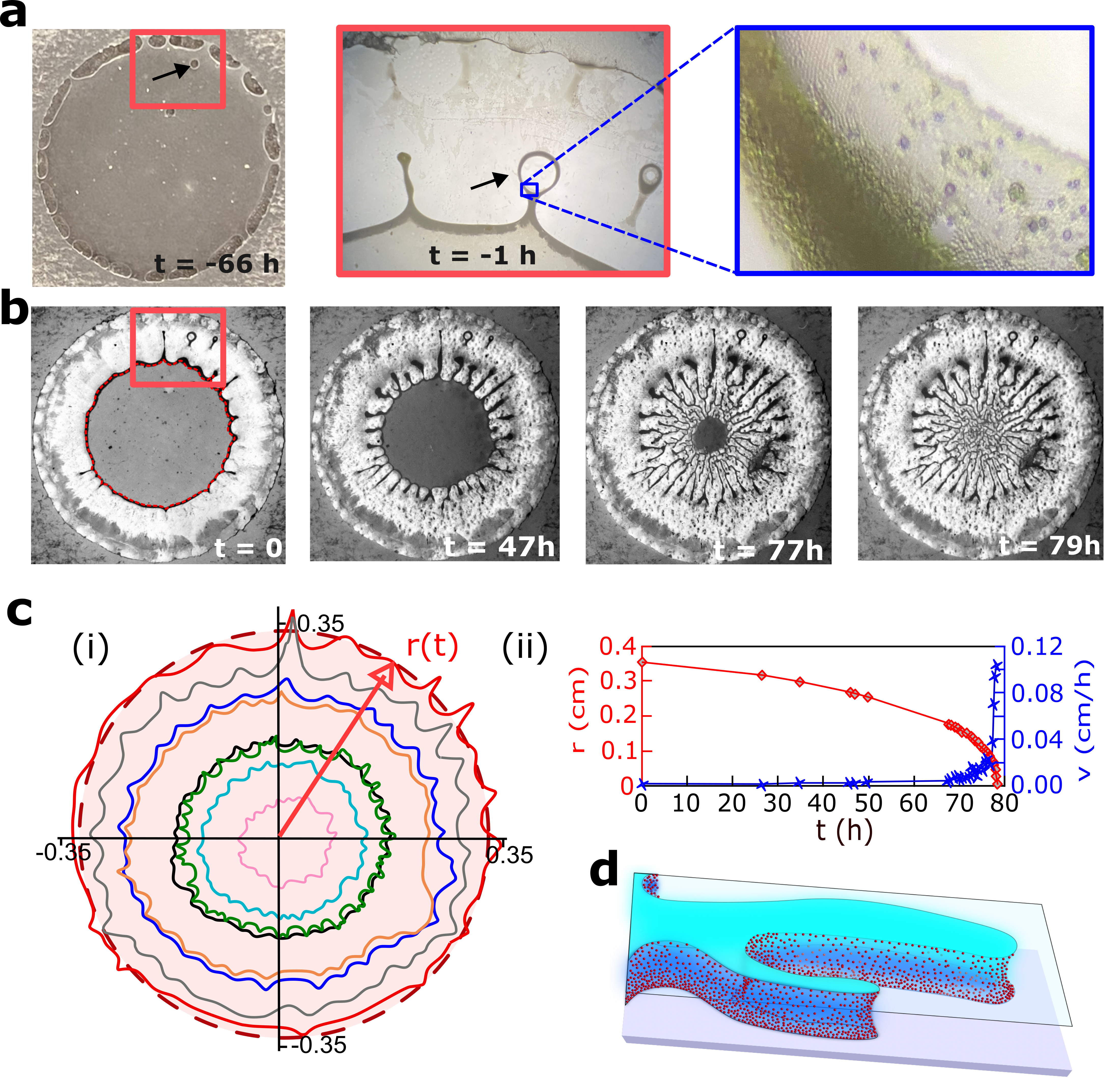}
    \caption{Drying kinetics of a confined droplet containing 1\,wt\% of fluorescently labelled, 1.8\,$\mu$m large TPM particles. (a) Photograph of the sample, taken 72 hours after preparation and 66 hours before starting the time-lapse video V4. The photo and microscope images (red and blue squares) were taken 1 hour before starting the V3 recording and capture the onset of finger formation. The black arrows show an air bubble that is trapped by the accumulation of colloids at the water-air interface. Other air bubbles escape to the outer rim of the sample. (b) Snapshots from the time-lapse series shown in video V4, starting at the relative time stamp $t = 0$. The sequence of photos shows the progressive shrinkage of the inscribed circle with average radius $r(t)$, excluding the fingers. (c(i)) Traces of the drop interface inscribing the area inside the black appearing rim of the droplets as a function of time. The outer red trace is taken for the photo taken at $t = 0$. (ii) Plot of the change of $r(t)$ (red) and velocity $v(t)$ (blue) of the inscribed circle. The solid lines are guides to the eye. (d) Illustration of individual fingers; they remain connected between the bottom and top surface until the entire pattern is formed.}
    \label{fig:4}
\end{figure}

The drying kinetics of a sample with radial symmetry is shown for different stages of pattern formation in Fig.\,\ref{fig:4}. The images in Fig.\,\ref{fig:4}b are taken from the time-lapse video V4. 

\textbf{Early stage}
The cylindrical cell that contains the suspension droplet has been filled, but not hermetically sealed.
Consequently, water evaporates and air flows in slowly from the outer rim of the cylindrical cell.
A visible manifestation of this throttled gas exchange is the slow growth of air bubbles along the outer rim of the cylindrical cell. 
These grow and merge to form a roughly circular air-vapor droplet boundary (capillary bridge).

In the experiments, this detached droplet dries in about 5 days from an initial radius of $\sim$5 mm.
The average observed drying speed is then of order 10$^{-6}$ cm/s, which is much slower than the estimated shrinking speed that the same droplet would have if it were in contact with dry air ($\sim$ 2.7 10$^{-5}$ cm s$^{-1}$ (see SI).
The fact that the observed shrinking speed is about thirty times slower than in dry air indicates that the relative humidity ($RH$) of the air outside the droplet is close to one. 
In fact, as
\[
\dot{r}_{\rm partial} = (1-RH) \dot{r}_{\rm sat} ,
\]
we estimate that the air around the water droplet is approximately $>$96\% saturated. 
At this relative humidity, water will form a very thin``wetting'' film on the glass (silica) surface outside the droplet. 
At a relative humidity of 96\%,  the height of this wetting layer is in the nanometer range,  which is much smaller than the size of the TPM particles. 
Hence, colloids will not enter this film spontaneously. 
However, there will be a flow through these thin films, dominated by the flux through their rims, which is well documented~\cite{dimitrov1996continuous,deegan2000contact,hu_larson2002}. The resulting high fluid-flow velocity (P\'{e}clet number appreciably larger than one) drags colloids from the droplet into the thin film (i.e. the films are connected to the central droplet; see Fig.\,S2). 

Initially, the droplet is roughly cylindrical. 
If the evaporation rate is controlled by the permeability of the tape bounding the cell, the amount of mass $M$ evaporating per unit time $\rho \mathrm{d}V/\mathrm{d}t$ is constant; $\rho$ is the density of water and $V = \pi r^2(t) h$ is the volume of the cylindrical droplet. 
The constant rate of water evaporation implies that, as long as the droplet remains cylindrical, the rate of change of the area of the droplet is $d r^2/d t = \mbox{\rm constant}$. 
Therefore, $r^2(t) = r_0^2 -\alpha t$, where $\alpha$ is a system-dependent constant.  
The experimental data for the initial droplet shrinkage agree well with the above function, even though we observe the deposition of a colloidal monolayer (\textbf{stage 2}, lasting from $t \sim -66$\,h until about $t \sim -1$\,h; see figure~\ref{fig:4}b and Fig.\,S3 - here we define t=0 as the moment shortly after the finger formation, stage 3, starts). 
After the fingering instability appears, we can still define the location of the shrinking front, but the effective radius of this front moves faster than the front of a shrinking cylindrical droplet. 

Due to the evaporative flux through the side of the droplet, charged TPM colloids will be advected to this interface (or, more precisely: the interface "sweeps up" colloids).
But the colloids can also diffuse.
The competition between the speed of advection ($v\approx 10^{-6}$cm s$^{-1}$) and the self-diffusion coefficients of TPM colloids in water($D$ = $\mathcal{O}(10^{-8})$ cm$^2$ s$^{-1}$ results in the formation of a fluid colloidal ``crust'' terminated by the air-water interface.
In steady state, when the number of colloids deposited balances the number of colloids advected, the typical thickness of this crust is $D/v$, which is of the order of dozens of colloidal diameters.
The above estimate yields just an order of magnitude because the ``sweeping up" of the colloids mainly takes place at the bottom of the cell: the colloids are then moved up due to the osmotic pressure of the dense suspension at the air-water interface, and diffuse away into the bulk of the droplet, where they sediment again.
However, a fraction of the colloids are swept along the air-water interface to the top rim of the droplet, where they are deposited in a monolayer on the top surface of the cell (Fig.\,\ref{fig:2}b(v)). 

\textbf{Monolayer to finger transition.}
As the central droplet shrinks, three things happen: (1) The area of the external, colloid-supported wetting film grows and carries a growing fraction of the evaporative flux~\cite{denkov1993two}. 
(2) The radial shrinkage speed of the droplet increases, as explained above.
(3) There is an accumulation of colloids near the rim of the droplets.

The result of these effects is that, at some point, the transport of fluid to the colloid-supported film cannot keep up with the rate of evaporation of this film.
When this happens, the film breaks, and the evaporative flux through the rims of the droplet drops sharply, causing the 
colloid deposition outside the droplet to stop, as seen in the experiments.
From then on, the droplet shrinks mainly due to evaporation through its vertical surfaces (\textbf{stage 3}).

\textbf{Surface instability.}
As the suspension inside the droplet is fluid, colloids are still being transported to the liquid-vapor interface by the evaporative flow. 
However, as the colloids are now no longer deposited in the wetting film, they accumulate on the vertical surface of the droplet, thickening the colloidal ``crust''  at the air-water interface (Fig.\,S2).
The flow of solvent causes an additional stress in the crust.
As the crust is fluid, it is plausible that this stress is isotropic and has therefore a transverse component that counteracts the surface tension of the air-water interface.
At some point, this transverse stress overcomes the air-water surface tension, and the interface starts to crumple (for a description of the buckling of ``solid'' particle crusts, see {\em e.g.}~\cite{Bamboriya2023})

Importantly, as the overall surface tension of the droplet is now negative, the Laplace pressure of the tips of the fingers is negative, whilst it is positive at the points of maximum indentation~\cite{eriksen2018pattern}.
The tips of the fingers will not move, because such motion would increase, rather than decrease the volume of liquid in the fingers. 
However, the water transport continues to feed the evaporation through the surface of the fingers.
Evaporation of the water in the fingers compacts the colloidal suspension in the fingers, to the extent that they undergo structural arrest.
Indeed, we observe in the experiments that fingers, once formed, retain their shape.

However, the tips of the air fingers can move inwards, as they face the much more dilute colloidal suspension in the remaining central part of the droplet.
As more water evaporates, the colloidal concentration in the central part of the droplet continues to increase, and at some point, the Laplace pressure of the invaginations is insufficient to move the surface.
At that point, the droplet undergoes a secondary fingering instability, such that curvature at the tip of the air fingers can overcome the colloidal osmotic pressure.
Indeed, in the experiments, we observe a fairly sudden splitting of the air fingers.
In principle, this scenario could be repeated, but in our experiments, it is observed only once, before the colloidal suspension inside the fingers undergoes structural arrest. 

\section{Conclusion}

The experiments presented here demonstrate that the slow evaporation of a confined, particle-laden droplet can be qualitatively different from the patterns that form during the drying of sessile droplets.

We note that it is the confinement that is the crucial factor in the emergence of the fingered patterns that we observe.
The fact that the droplet is cylindrical is not essential.
In fact, similar patterns form when studying the drying of a colloidal suspension in a rectangular cell (see Fig.\,S5).

As multilayer structures are common in natural materials, it seems likely that the kind of patterns that we observe in confined cells, also occur in naturally occurring materials. 

\section{Methods}

\subsection{Chemicals}

All chemicals purchased were used as received. 3-(trimethoxysilyl)propyl methacrylate (TPM) oil (98\%, Sigma Aldrich), ammonium hydroxide solution (28\% NH3 in H20, 99.99\%, Sigma Aldrich), 2,2-Azobis(2-methylpropionitrile) (AIBN 98\%, Sigma-Aldrich), Emprove Expert 4M hydrochloric acid (HCl, Sigma Aldrich), DMSO (ACS reagent >99\%, Sigma Aldrich), THF, BODIPY 505/515 (Thermo Fischer Scientific), Chloroform (anhydrous, 99\%, Sigma Aldrich), 3-aminopropyl trimethoxysilane (97\% Sigma Aldrich), Pluronic F-127 (Sigma Aldrich), and deionised water from Merck Millipore Milli-Q Direct-Q 3 UV purification system. 

\subsection{Synthesis}

We synthesised TPM (3-(trimethoxysilyl)propyl methacrylate) particles of diameter $d_{p} \sim 1.8$\,$\mu$m, following a simple bench-top synthesis route\,\cite{van2017preparation}. For this we measured 100\,mL deionised water in a 250\,mL beaker. While stirring the water on a magnetic stirrer (400rpm) 100\,$\mu$L of 28\,wt.\% ammonia was added, followed by the addition of 1\,mL TPM oil. We then allow an emulsion to form for 1hr and 15 minutes. Subsequently, we added 10\,mg of AIBN (azobis(isobutyronitrile)), a radical initiator. After a couple of minutes, the mixture was transferred to a flask and sealed with a screw top, and heated at 80\,$^{\circ}$C for 3 hours. The resulting colloidal dispersion was washed with deionized water in 3 centrifuging and decanting cycles. The zeta potential of these particles is reported to be about -36\,mV at pH 5.6, their density is 1.314\,g\,cm$^{-3}$, and their refractive index is 1.512\,\cite{van2017preparation}.

For use in confocal microscopy, we used a different synthesis so as to incorporate a fluorescent dye into the particles\,\cite{liu2019colloidal}. This method produced particles with a slightly smaller diameter of $\sim 1.3$\,$\mu$m. To do this, we first added 2mg of BODIPY to 2ml of chloroform with 2$\mu$l of 3-aminopropyl ethoxysilane and leave to react for 24 hours. The following day we started by mixing 20ml of 0.5mM HCl solution with 2ml TPM oil for 1 hour until mixture is no longer cloudy. To make 1.3$\mu$m diameter particles, we mixed 6ml of the resulting solution with 19ml of 0.5mM HCl solution in a 100ml round bottom flask for 10 min (600rpm). Next, we added 25ml of 0.028\% ammonia solution and let it sit for 45 minutes. Then 400$\mu$l of the BODIPY solution was added and mixed for 1 hour, and following this 10mg of AIBN was added and the mixture was left to stir for another 30 minutes. Finally, the mixture was placed in an oven at 80\,$^{\circ}$C and tumbled every 30 minutes for 3 hours. The resulting suspension was cleaned in the same way as before. 

In either case, the final TPM suspension was kept for later use at 4\,$^{\circ}$C.

\subsection{Bright- and Dark-Field Microscopy and Photography}

Bright-field and dark-field microscopy were performed using a Amac microscope, equipped with a 5x, 10x and 40x objectives, and a CMOS camera. Larger fields of view were photographed with an i-phone 13 through the eye-piece with 10x magnification. Timelapse movies were recorded with a FUJIFILM X100F digital camera with a Fujinon aspherical lens, Super EBC, $f=23$\,mm 1:2, from Japan.

\subsection{Confocal Microscopy}

The TPM particles were fluorescently labelled, either during the synthesis process, as described before, or by following a swelling-deswelling procedure, after the basic synthesis, introduced by Oh et al.\,\cite{oh2015high} and described for polystyrene particles by Zupkauskas et al.\,\cite{zupkauskas2017optically}: We mixed 10\,mL of a 1wt\% TPM suspension with 20\,mL of THF (tetrahydrofuran). Subsequently, 200\,$\mu$L of BODIPY in DMSO (1\,mg/ml), and the mixture was rigorously stirred for 30\,min. Then an excess of 200\,mL DI-water, heated at 60\,$^{\circ}$C for 20\,min, and the washed 3x by centrifugation.
A Leica DMi8 SP8 confocal microscopy was used to image the drying of the confined TPM suspension.

\begin{acknowledgments}
The authors acknowledge the support of the Research Council of Norway through its Centres of Excellence funding scheme, project number RCN 262644. E.E. acknowledges funding from the European Union's Horizon Europe research and innovation programme Fluxionic under the Marie Skłodowska-Curie grant agreement No 674979. The authors thank Daan Frenkel for extended input in discussion concerning the theory, and manuscript writing. E.E. thanks Michael Cates, Benjamin Rotenberg and Simone Meloni for their comments and discussions.
\end{acknowledgments}

\section{References}
\bibliography{conf-coll-circ-maze}

\appendix

\renewcommand{\thefigure}{S\arabic{figure}}
\setcounter{figure}{0}

\section{Supplementary Figures}

\begin{figure}
    \centering
    \includegraphics[scale=0.5]{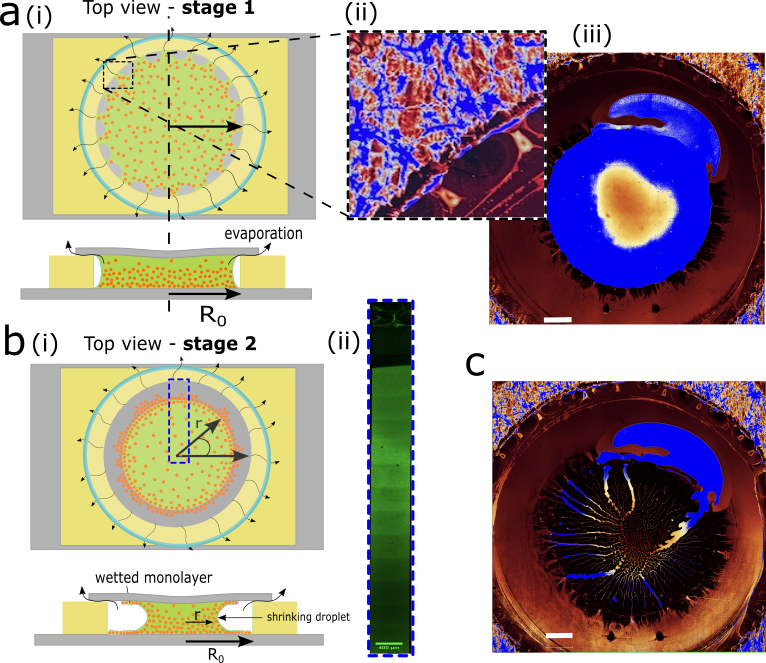}
    \caption{Schematic and confocal images of a confined drying droplet containing 1.8\,$\mu$m large TPM particles, and fluorescently dyed water, shown at different drying stages. \textbf{a}\,(i) Schematics of the top-down and cross-sectional view of the sample at \textbf{stage 1} of the drying process, showing the formation of air bubbles at the interface between the liquid and the double sided tape. (ii) Confocal image of the top surface of the cell that shows the fluorescent water trapped between the coverslip and double sided tape spacer. Blue indicates a high, and red a very low, water content. The almost black regions bounded by the red to yellow inside the bubbles are dry and almost void of colloids. (iii) 5$\times$5 tiled confocal image show the entire sample cell (total image size 2560$\times$2560 pixels), taken at the onset of \textbf{stage 3}. The scale bar is 1\,mm. The schematics in \textbf{b}(i) illustrates the onset of monolayer deposition after the air bubbles have coalesced. The blue dashed rectangle corresponds to the composite confocal-image stacks in (ii), taken from the centre of the sample to the tape at the edge of the cell. The sharp black-to-green interface at the top, shows the outer air-suspension interface, while the integrated intensity decreases toward the centre of the sample, reflecting a dimple in the coverslip cause by capillary pressure. \textbf{c} A tiled confocal image of the sample (after 4 hours and 20 minutes relative to \textbf{a}\,(iii)) when the entire pattern has formed, but is still wet. See the full drying sequence in Video \textbf{V2}.}
    \label{fig:S1}
\end{figure}

Fig.\,\ref{fig:S1}\,\textbf{a}(i) illustrates schematically (with a top-down and cross-sectional view) how the water slowly evaporates through the gap between the coverslip and the double sided tape that serves as spacer and confining walls of the flat sample cell containing 1\,wt\% of 1.8\,$\mu$m large TPM particles. 
Here, the water was fluorescently labelled with the green fluorescent dye calcein (excitation at 495\,nm, emission at 515\,nm) in order to visualize the water in the sample. 
Fluorescent images (512$\times$512 pixels, with pixel size 4.55\,$\mu$m) were taken with a 5$\times$ objective lens, while focused in the middle of the cell (approximately 40\,$\mu$m from the bottom surface), and with a long exposure time to visualise where, and how much, of the fluorescent water is present. In all images, blue indicates a high water content (saturating the monitor for the green dye), red indicates a much lower water content and yellow an intermediate one.

Fig.\,\ref{fig:S1}\,\textbf{a}(ii) shows a zoomed in image taken at the edge of the cell that demonstrates that water and air can be transported between the coverslip and the double sided tape. Fig.\,\ref{fig:S1}\,\textbf{a}(iii) is a 5$\times$5 grid of images taken, as described above, to generate a picture of the entire sample at the moment finger formation starts. We refer to this drying period as \textbf{stage 3} in the main text. 
This stage signifies the moment when the monolayer deposition on top and bottom surface stops and fingers start to form. The yellow to red coloring in the center of the droplet reflects the depression of the coverslip also depicted schematically in the sample cross-section in \textbf{a}(i) and \textbf{b}(i).
This depression appears within the first 6 hours of evaporation and remains until the entire fingering pattern has formed. This is due to the capillary force produced by the capillary bridge\,\cite{gennes2004}.
Assuming a wetting angle $\theta$ of roughly 45$^{\circ}$, an average height $h = 60$\,$\mu$m and radius of $r = 5$mm, we estimate the capillary force pushing the two surfaces together to be of the order of 0.15\,N, which is sufficient to deform the coverslip while the double sided tape represents the counteracting force.

Fig.\,\ref{fig:S1}\,\textbf{b}(ii) is a tiling of confocal image stacks taken across the sample thickness (50 images taken over a sample thickness of 80\,$\mu$m), shortly before the onset of finger formation. 
The entire strip is 7\,mm long and 0.6\,mm wide, starting from the middle of the sample (bottom) and ending at the tape at the sample edge (top). The drop of fluorescent intensity towards the center of the droplet (decreasing green levels) shows the depression of the sample due to capillary forces in the middle.  

Fig.\,\ref{fig:S1}\,\textbf{c} shows the tiling of confocal-images of the entire sample, in exactly the same way as for \textbf{a}(iii), but just after the fingering pattern had finished forming. The blue, yellow and red show that the fingers and the monolayers are still wet, but that the monolayers and central fingers dry first while the finger tips dry last. The entire sequence of the finger formation is shown in Supporting Video \textbf{V2}, in which the individual images were taken in 20 minute intervals, over the course of 4 hours and 20 minutes. 

\begin{figure}
    \centering
    \includegraphics[scale=0.5]{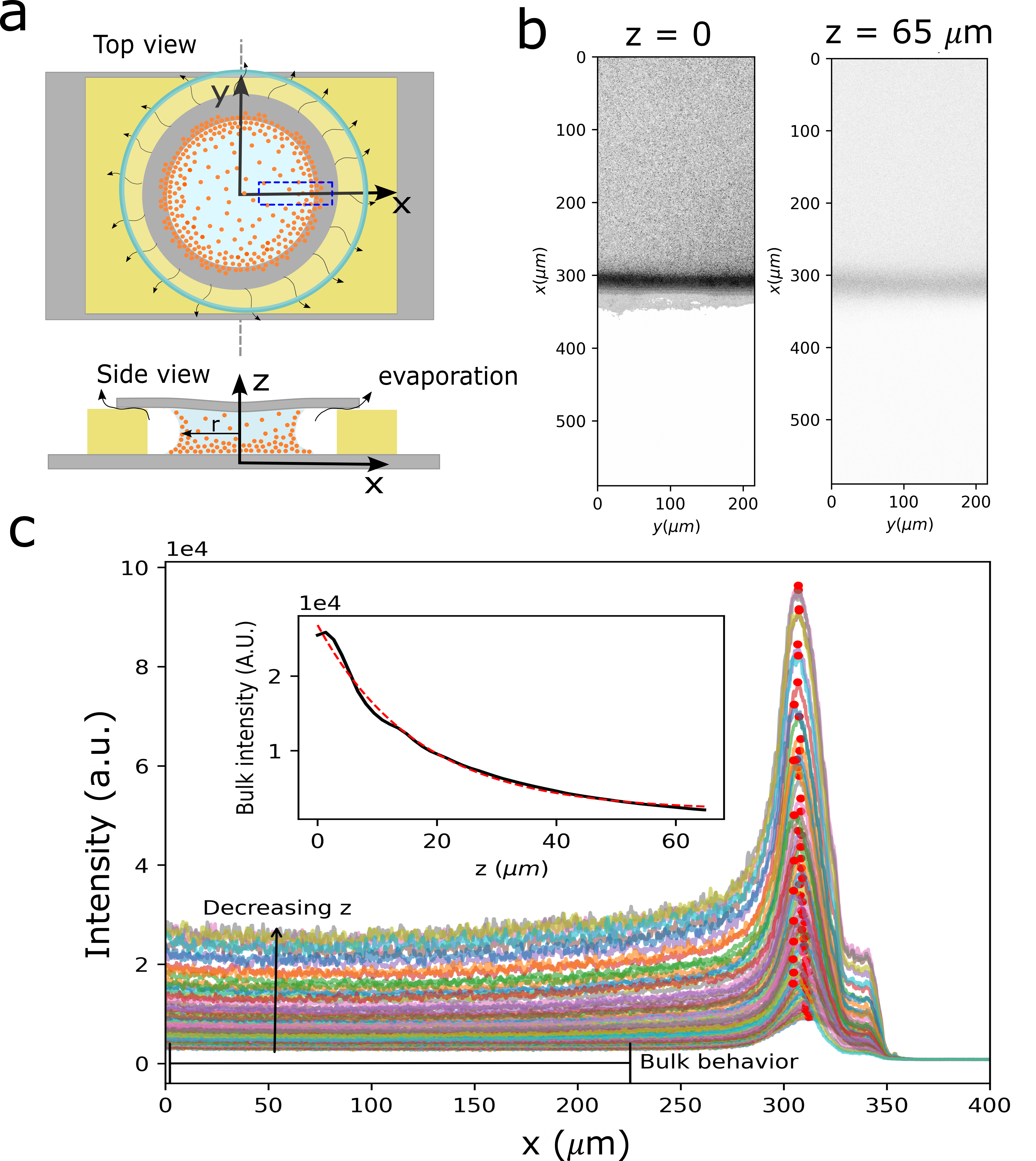}
    \caption{Colloid distribution at the droplets' air-water interface. \textbf{a}\, Schematic of a droplet of the colloidal suspension placed in the middle of a sample cell, with both top-down and cross-sectional views. As this was placed in the middle, rather than allowed to dry to this stage on its own, monolayer deposition has only just begun at the droplet edge at this stage. The blue dotted rectangle corresponds to the area shown in the two confocal images in \textbf{b}. The two images are individual slices from a stack on confocal images through the entire thickness of the sample (approx. 80\,$\mu$m). The image on the left hand side is taken at z = 0, corresponding to the bottom surface of the cell, while the image on the left is taken at z = 65 microns (approx. 15\,$\mu$m below the top surface). \textbf{c}\, An intensity plot of all the images within the confocal image stack of the interface, stacked in the z direction from top to bottom. The inset plot shows the average bulk intensity from the bottom of the cell (z = 0) almost to the top of the cell at z = 65\,$\mu$m.}
    \label{fig:S2}
\end{figure}

In Fig.\,\ref{fig:S2}\,\textbf{a} we illustrate schematically both a top-down and cross-sectional view of a small droplet of the colloidal suspension placed in the middle of the sample cell, such that it does not reach the wall of the cell made by the double sided tape. This allowed us to observe the interface between the vapor phase and the droplet confined between the top and bottom surface in a controlled way. Here we used 1.3\,$\mu$m large TPM particles that were fluorescently labelled with BODIPY\,\cite{liu2019colloidal}.

Fig.\,\ref{fig:S2}\,\textbf{b} shows two confocal microscopy images taken across a 200\,$\mu$m wide section at the air-droplet interface (blue dashed rectangle in the top view); they were recorded at the bottom ($z$ = 0) and near the top ($z$ = 65\,$\mu$m) of the sample and are part of a $z$-stack of images leafing through the entire sample thickness, which is shown in Supporting Video \textbf{V3}. In Fig.\,\ref{fig:S2}\,\textbf{c} we plot the total intensity obtained for each $z$-layer of 2\,$\mu$m thickness and averaged in $y$-direction that are shown in \textbf{V3}. 
The grey-scales in Fig.\,\ref{fig:S2}\,\textbf{b} indicate the intensity of the fluorescence coming form the colloids: higher particle concentrations show stronger scattering intensity and thus a darker gray. The outside of the droplet is white as it is completely void of particles. At $z =0$\,$\mu$m, we see the onset of a colloidal monolayer deposition at around $x \approx 340$\,$\mu$m followed by a sharp intensity increase at $x \approx 310$\,$\mu$m (thick dark line) that coincides with the onset of the steep vapour-droplet interface. At even smaller $x$-values the scattering intensity levels off, reflecting a constant bulk concentration of the fluorescently labelled colloids inside the droplet. From the intensity curves across the vapor-droplet interface we estimated the mean-width half-value (MWHV) as function of $z$. We found a MWHV of roughly 25\,$\mu$m for all $z$-values. However, the colloid density of the particles at the bottom is higher than at the top.
 
In Fig.\,\ref{fig:S2}\,\textbf{c} the intensity curves obtained from integrating the intensity (gray scale) in the $xy$-plane are plotted as function of $z$ in a staggered way so that the intensity at the top-surface appears at lower intensity values and the bottom surface is shown at higher intensity values. In the inset we plot the intensity values as function of $z$ for each intensity curve, summed up from $x$ = 0 to 225\,$\mu$m). The dashed line is an exponential fit to the intensity curve showing that near the air-water interface the colloids are Boltzmann distributed in $z$.

\begin{figure}
    \centering
    \includegraphics[scale=0.6]{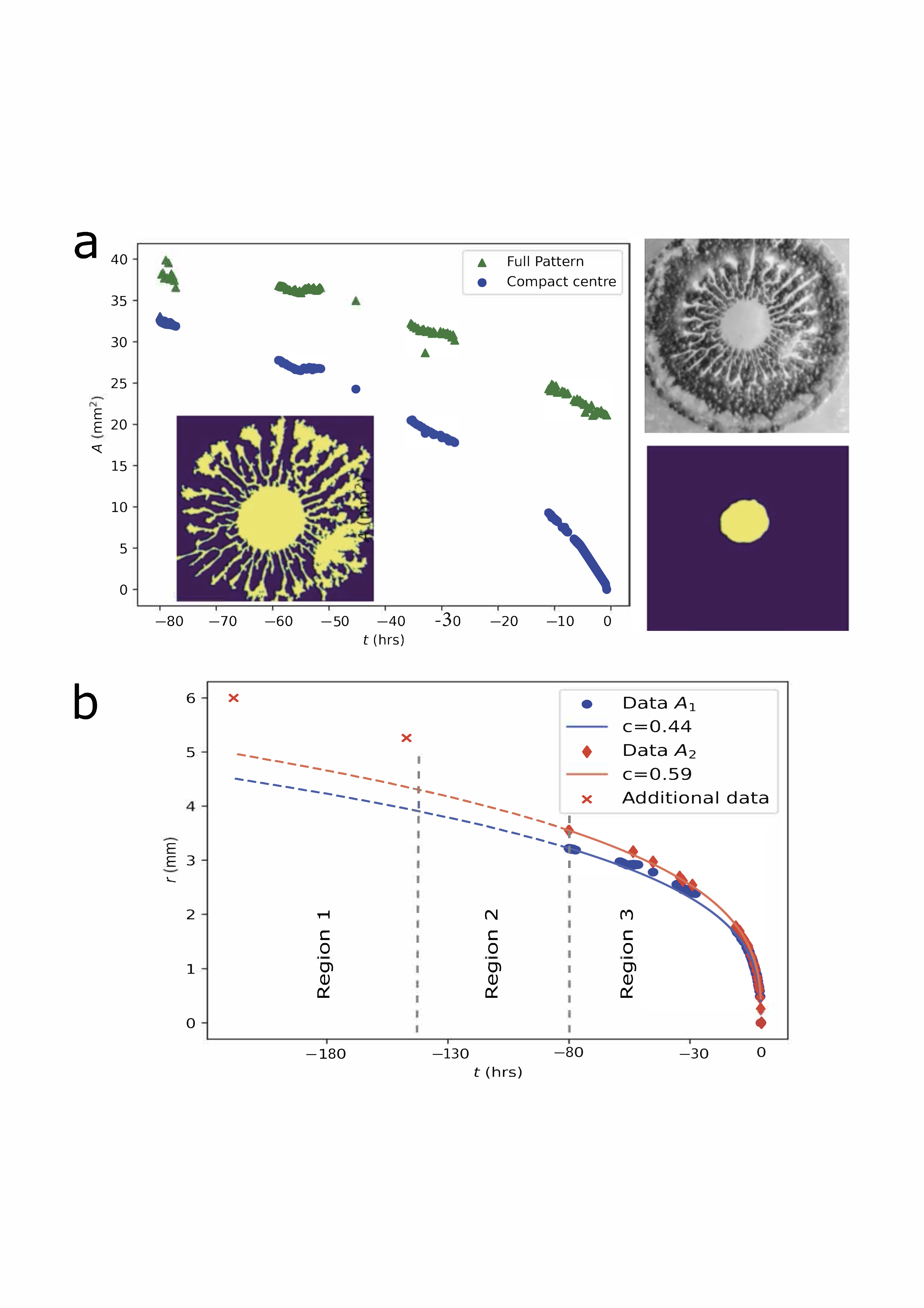}
    \caption{Analysis of the drying kinetics shown in Fig.\,3 in the main text. \textbf{a} Plot of the inscribed area $A_1$, which was estimated using method 1 (compact centre) and the area of the full pattern that includes also the area of the fingers. Snapshots of the respective thresholded patterns obtained for a given time at around $t=70$\,hours are shown, where the area is yellow and the background is purple. The corresponding original photograph is also shown. \textbf{b} Average radius of the inscribed circle extracted from the areas of the inscribed circles plotted in \textbf{a} and using method 2, which correspond to the data for $r(t)$ shown in Fig.\,3 in the main text. The corresponding fitting lines are plotted as solid lines and continued to the start of the drying process at $t=-166$\,hours as dashed lines. Region 1 corresponds to the bubble formation and coalaescence, region 2 to the monolayer deposition and region 3 to the finger formation.}
    \label{fig:S3}
\end{figure}

In Fig.\,\ref{fig:S3} we present different ways of analysing the shrinking area of the confined droplet as function of time, which is presented in Fig.\,3 in the main text and video \textbf{V4}. Once finger deposition starts, the progressively shrinking drop area $A$, as seen from the top, can be described either by the total area of the fingers plus the inscribed circle or by the inscribed circle only. We developed a Python code to measure these two areas from the time-lapse video \textbf{V4}.

In the analysis, we make use of two different methods of calculating the inscribed circle. Method A1 is based on applying a Sobel filter on the time-lapse images to find the lines of maximum contrast in the images. Then, we apply a Gaussian blur to be able to distinguish the centre from the fingers. We perform an image threshold based on Otsu's algorithm to binarise the images, and then we perform image segmentation based on the Hoshen-Kopelman algorithm to separate all colloid-rich (i.e. whiter) areas of the images. Subsequently, we count the number of pixels in the central cluster, which is the total area of the inscribed circle $A_1$. We analyse the radius of this circle, which we denote $r_1(t)=r_1$.

Fig.\,\ref{fig:S3} also contain two smooth curves, which are based on the scaling relation
\begin{equation}\label{eq: Scaling Relation R1}
     \frac{dr_1}{dt} \sim 1/r_1^2 \iff \frac{dA_1}{dt} \sim 1/r_1
\end{equation}
Assuming that $r_1\rightarrow0$ at the end of the drying process, we find the proportionality constant in eq. \eqref{eq: Scaling Relation R1} to be 
$$C = -\frac{1}{3\Delta t}r_{1,0}^3\ ,$$
where $r_{1,0} \approx 3.5$\,mm is the radius of the inscribed circle when we start to observe the fingering instability, and $\Delta t$ is the elapsed time of the instability (i.e. from $t=0$ in Fig.\,\ref{fig:S3}). 

In the second method, we assessed the area $A_2(t)$ using the WebPlotDigitizer from Automeris LLC. In this method we estimated a slightly larger area than for method 1, as it considered the outer perimeter of the inscribed circle but excluded the finger. We could fit the resulting area $A_2(t)$ with the same functional form with a larger constant ($C=0.59$).

We observe that this scaling relation only applies \emph{after} the fingering instability occurs, as seen by the extrapolated dashed lines in Fig.\,\ref{fig:S3}. Thus, we have a transition from an expected linear scaling $r^2(t) = r_0^2 -\alpha t$ with $\alpha$ as the system-dependent drying rate, to the scaling relation proposed for the fingering region.

\begin{figure}
    \centering
    \includegraphics[scale=0.6]{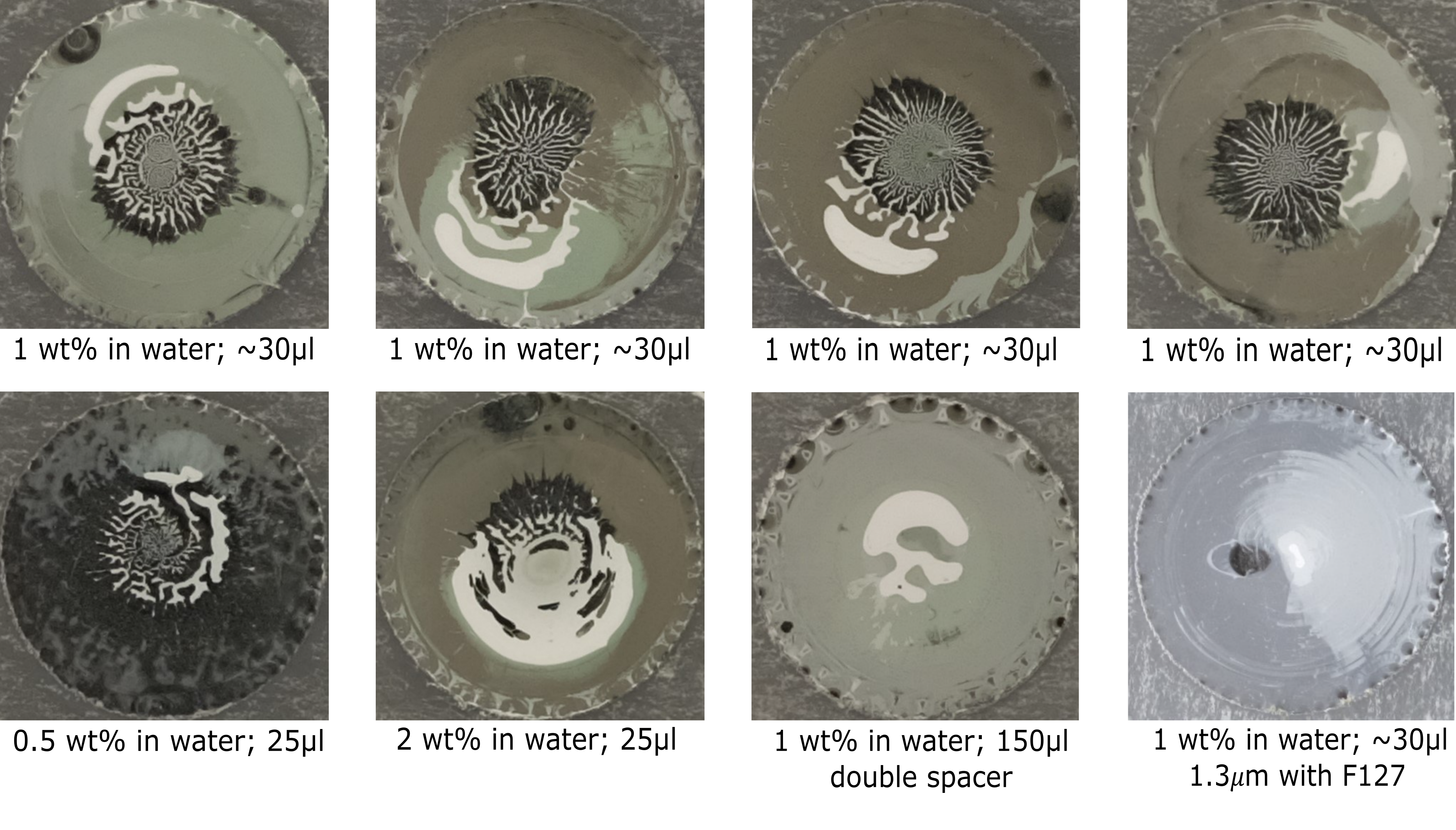}
    \caption{Gallery of samples photographed in reflection. The top row shows 4 other samples with the same composition as those in the main text, showing high levels of reproducibility. The bottom row shows some other samples, each with a different modification. From left to right: 0.5wt\% concentration of TPM, 2wt\% concentration of TPM, double-spaced surfaces (cell height of 160\,$\mu$m), and 1.3\,$\mu$m large TPM colloids that have been surface functionalised with Pluronic\textregistered\, F127.}
    \label{fig:S4}
\end{figure}

\begin{figure}
    \centering
    \includegraphics[scale=0.6]{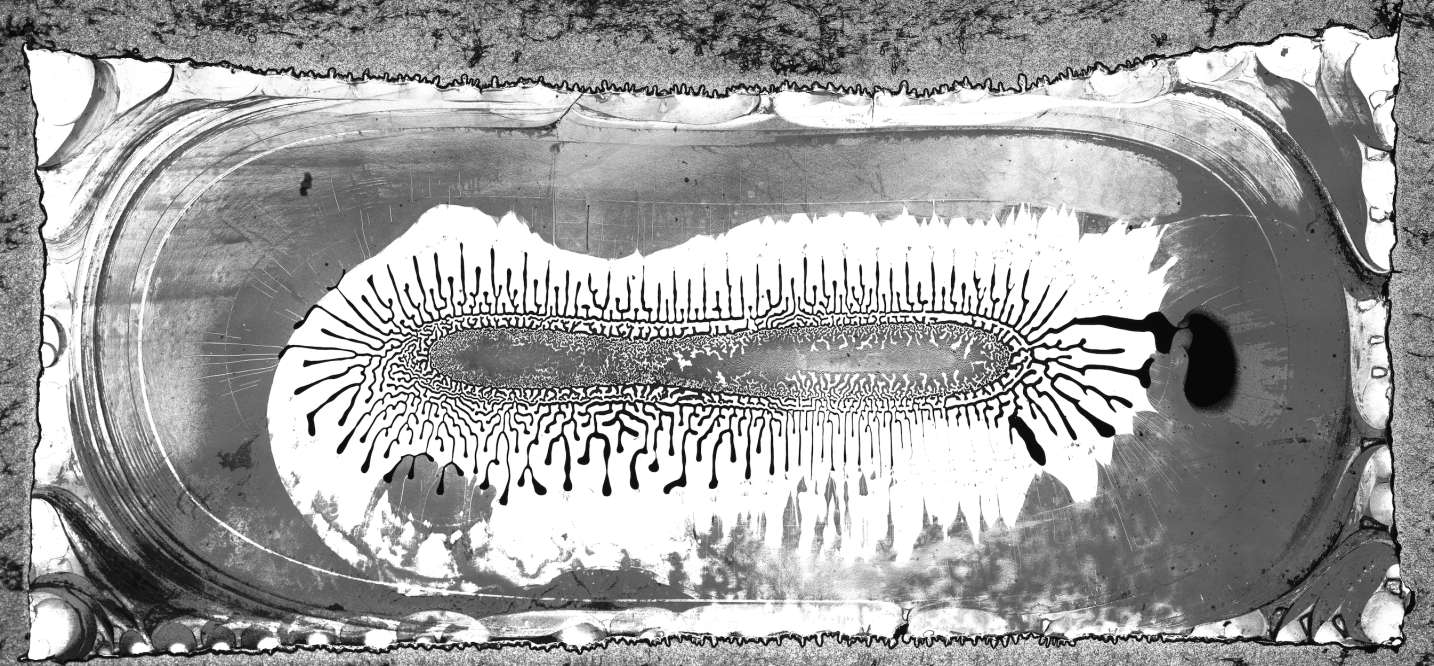}
    \caption{Transition micrograph of a dried sample that was prepared with a suspension of 1\,wt\% of 1.8\,$\mu$m large TPM particles. The image was reconstructed from a 12$\times$6 tiling of micrograpghs. The square cell is 24 $\times$ 12 mm large.}
    \label{fig:S5}
\end{figure}

In the first row of Fig.\,\ref{fig:S4}, we show the final drying patterns of several other samples, made in the same way as those shown in the main text, to demonstrate the highly reproducible nature of these patterns. In the second row, we then show 4 more samples, each prepared under slightly different conditions. Typically, using lower concentrations of TPM particles lead to similar drying patterns, but with a less dense monolayer deposition than the samples containing 1\,wt\% TPM particles. In contrast, using higher concentrations of TPM particles lead to similar monolayer deposition behaviour as for the samples containing 1\,wt\% TPM particles, but produced much less clearly defined patterns in the centre. Using a double spacer with a cell height of $\sim 160\,\mu$m lead to the onset of thick fingers that were kinetically hindered to form the transition to finer fingers at the centers, but we still observe monolayer deposition as before. Adding the polymeric triblock surfactant Pluronic\textregistered\, F127 suppressed finger formation completely.
In Fig.\,\ref{fig:S5}, we also show that we observed the same pattern formation phenomenon with a rectangular geometry.

\section{Rate of evaporation}

In the experiments, the detached droplet in Fig.\,4 dries in about 5 days from an initial radius of 5mm.
The average observed drying speed is then of order 10$^{-6}$\,cm\,s$^{-1}$, which is much slower than the estimated shrinking speed that the same droplet would have if it were in contact with dry air ($\sim$ 2.7 $\times$ 10$^{-5}$\,cm\,s$^{-1}$). 

We can estimate how quickly a cylindrical droplet of the same size would evaporate into dry air at 25$^{\circ}$\,C. 
The steady-state concentration profile around a cylindrical droplet with radius $R$ in a cell with a radius $R_0$ depends logarithmically on the distance $r$ from the center of the droplet:
\[
\rho(r)=\rho(R) +\frac{\ln{r/R}}{\ln{R_0/R}}(\rho(R_0)-\rho(R)) ,
\]
where $\rho(r)$ denotes the number density of the water vapor at radius $r$. 
We assume that the density of the water vapor at the boundary of the droplet is that of the saturated vapor $\rho(R)=\rho_{\rm sat}$, where $\rho_{\rm sat}$ is the saturated water vapor pressure at ambient conditions. 
If the cylindrical droplet is evaporating into almost dry air, we can assume that the transport of water vapor is diffusion-controlled.
The radial concentration gradient is then:
\[
\nabla c(r) = -\rho_{\rm sat} \frac{1-R_H}{r\ln{R_0/R}}.
\]
where $R_H$ is the relative humidity of the air at the outer boundary of the cell. 
The total diffusive flux of water vapor per unit height $J_D$, is
\[
J_D = \frac{2\pi D \rho_{\rm sat}(1-R_H)}{\ln{R_0/R}}.
\]
The diffusion coefficient of water in air at 25$^{\circ}$\,C is approximately 0.24 cm$^{2}$\,s$^{-1}$.
As the droplet shrinks, $R_0/R$  is not constant, but to obtain an estimate, we take a typical value:  $R=0.5 R_0$.
The flux of water (per unit length) that must evaporate due to the shrinkage of the droplet equal to
\[
J_s = -\rho_L \pi \frac{d r^2}{d t} .
\]
Balancing the two fluxes, we get:
\[
 \frac{2 D \rho_{\rm sat}(1-R_H)}{r\ln{R_0/R}}= 2 \rho_L \dot{r} 
\]
The saturated vapor pressure of water vapor at 25$^{\circ}$\,C is $\approx$ 0.032 atm. 
Hence, its molar volume is 7.8 $\times$ 10$^5$cm$^3$ (assuming an ideal gas molar volume of 24.46 l).
Under the same conditions, the molar volume of liquid water is approximately 18 cm$^3$. 
Therefore, if $R_H=0$ (dry air):
\[
\dot{r} \approx  \frac{ D \rho_{\rm sat}/\rho_L}{r\ln{R_0/R}} \approx 2.7 10^{-5} \mathrm{cm}\,\mathrm{s}^{-1}.
\]
If the water is in contact with air with a relative humidity RH, then the rate of evaporation is reduced, and the shrinkage of the droplet slows down
\[
\dot{r}_{\rm partial} = (1-RH) \dot{r}_{\rm sat} .
\]
As the experimentally observed shrinking speed is of the order 10$^{-6}$\,cm\,s$^{-1}$, we conclude that the relative humidity of the air at the outer rim of the cylindrical cell is approximately 96\%.

As the atmosphere outside the cell has a typical relative humidity of $\sim$30-50\% (Norway), it follows that largest drop of the chemical potential of the water is inside the scotch-tape seal.
The implication is that the water evaporative flux is effectively constant until the droplet has disappeared.

\end{document}